%% file: lrec.tex
\documentclass[10pt, a4paper]{article}
\usepackage{lrec}
\usepackage{multibib}
\newcites{languageresource}{Language Resources}
\usepackage{graphicx}
\usepackage{tabularx}
\usepackage{soul}
\usepackage{titlesec}
\titleformat{\section}{\normalfont\large\bf\center}{\thesection.}{1em}{}
\titleformat{\subsection}{\normalfont\SmallTitleFont\bf\raggedright}{\thesubsection.}{1em}{}
\titleformat{\subsubsection}{\normalfont\normalsize\bf\raggedright}{\thesubsubsection.}{1em}{}
\renewcommand\thesection{\arabic{section}}
\renewcommand\thesubsection{\thesection.\arabic{subsection}}
\renewcommand\thesubsubsection{\thesubsection.\arabic{subsubsection}}
\usepackage{placeins}
\usepackage{comment}
\usepackage{multirow}

\usepackage{epstopdf}
\usepackage[utf8]{inputenc}

\usepackage{hyperref}
\usepackage{xstring}

\usepackage{color}

\title{The STEM-ECR Dataset: Grounding Scientific Entity References in STEM Scholarly Content to Authoritative Encyclopedic and Lexicographic Sources}

\name{Jennifer D'Souza, Anett Hoppe, Arthur Brack, Mohamad Yaser Jaradeh, S\"oren Auer, Ralph Ewerth}

\address{TIB Leibniz Information Centre for Science and Technology, \\
         Hannover, Germany \\
         \{jennifer.dsouza,anett.hoppe,arthur.brack,yaser.jaradeh,auer,ralph.ewerth\}@tib.eu\\}

\abstract{
We introduce the STEM (Science, Technology, Engineering, and Medicine) Dataset for Scientific Entity Extraction, Classification, and Resolution, version 1.0 (STEM-ECR v1.0). The STEM-ECR v1.0 dataset has been developed to provide a benchmark for the evaluation of scientific entity extraction, classification, and resolution tasks in a domain-independent fashion. It comprises abstracts in 10 STEM disciplines that were found to be the most prolific ones on a major publishing platform. We describe the creation of such a multidisciplinary corpus and highlight the obtained findings in terms of the following features: 1) a generic conceptual formalism for scientific entities in a multidisciplinary scientific context; 2) the feasibility of the domain-independent human annotation of scientific entities under such a generic formalism; 3) a performance benchmark obtainable for automatic extraction of multidisciplinary scientific entities using BERT-based neural models; 4) a delineated 3-step entity resolution procedure for human annotation of the scientific entities via encyclopedic entity linking and lexicographic word sense disambiguation; and 5) human evaluations of Babelfy returned encyclopedic links and lexicographic senses for our entities. Our findings cumulatively indicate that human annotation and automatic learning of multidisciplinary scientific concepts as well as their semantic disambiguation in a wide-ranging setting as STEM is reasonable.
\\ \newline  }

\begin{document}

\maketitleabstract

\section{Introduction}
\input{introduction}

\section{Related Work}
\input{related_work}

\section{Scientific Entity Annotations}
By starting with a STEM corpus of scholarly abstracts for annotating with scientific entities, we differ from existing work addressing this task since we are going beyond the domain restriction that so far seems to encompass scientific IE. For entity annotations, we rely on existing scientific concept formalisms~\cite{Liakata2010CorporaFT,constantin2016document,augenstein2017semeval} that appear to propose generic scientific concept types that can bridge the domains we consider, thereby offering a uniform entity selection framework. In the following subsections, we describe our annotation task in detail, after which we conclude with benchmark results.

\begin{table}[!ht]
\small
\begin{tabular}{|p{7.5cm}|}
\textsc{\textbf{Process}} Natural phenomenon, or independent/dependent activities. E.g., growing (\textit{Bio}), cured (\textit{MS}), flooding (\textit{ES}).                                          \\
\textsc{\textbf{Method}} A commonly used procedure that acts on entities. E.g., powder X-ray (\textit{Che}), the PRAM analysis (\textit{CS}), magnetoencephalography (\textit{Med}).                                  \\
\textsc{\textbf{Material}} A physical or digital entity used for scientific experiments. E.g., soil (\textit{Agr}), the moon (\textit{Ast}), Doppler lidars (\textit{Eng}).                                                   \\
\textsc{\textbf{Data}} The data themselves, or quantitative or qualitative characteristics of entities. E.g., rotational energy (\textit{Eng}), tensile strength (\textit{MS}), the Sylow p-groups (\textit{Mat}).
\end{tabular}
\caption{The four scientific concepts studied}
\label{table:1}
\end{table}

\subsection{Our Annotation Process}
The corpus for computing inter-annotator agreement was annotated by two postdoctoral researchers in Computer Science.\footnote{We use BRAT~\cite{stenetorp2012brat} for annotating entity spans and their categories.} To develop annotation guidelines, a small pilot annotation exercise was performed on 10 abstracts (one per domain) with a set of surmised generically applicable scientific concepts such as \textsc{Task}, \textsc{Process}, \textsc{Material}, \textsc{Object}, \textsc{Method}, \textsc{Data}, \textsc{Model}, \textsc{Results}, etc., taken from existing work. Over the course of three annotation trials, these concepts were iteratively pruned where concepts that did not cover all domains were dropped, resulting in four finalized concepts, viz. \textsc{Process}, \textsc{Method}, \textsc{Material}, and \textsc{Data} as our resultant set of \textit{generic} scientific concepts (see Table~\ref{table:1} for their definitions).\footnote{We do not consider nested span concepts in this study, hence we leave out \textsc{Task}, \textsc{Object}, and \textsc{Results} since they were almost always nested with the other scientific entities. However, we found them as valid \textit{generic} categories as well.} The subsequent annotation task entailed linguistic considerations for the precise selection of entities as one of the four scientific concepts based on their part-of-speech tag or phrase type. \textsc{Process} entities were verbs (e.g., ``prune'' in \textit{Agr}), verb phrases (e.g., ``integrating results'' in \textit{Mat}), or noun phrases (e.g. ``this transport process'' in \textit{Bio}); \textsc{Method} entities comprised noun phrases containing phrase endings such as simulation, method, algorithm, scheme, technique, system, etc.; \textsc{Material} were nouns or noun phrases (e.g., ``forest trees'' in \textit{Agr}, ``electrons'' in \textit{Ast} or \textit{Che}, ``tephra'' in \textit{ES}); and majority of the \textsc{Data} entities were numbers otherwise noun phrases (e.g., ``(2.5$\pm$1.5)kms$^{-1}$'' representing a velocity value in \textit{Ast}, ``plant available P status'' in \textit{Agr}). Summarily, the resulting annotation guidelines hinged upon the following five considerations:

\begin{enumerate}
\item To ensure consistent scientific entity spans, entities were annotated as definite noun phrases where possible. In later stages, the extraneous determiners and articles could be dropped as deemed appropriate.

\item Coreferring lexical units for scientific entities in the context of a single abstract were annotated with the same concept type.

\item Quantifiable lexical units such as numbers (e.g., years 1999, measurements 4km) or even as phrases (e.g., vascular risk) were annotated as \textsc{Data}.

\item Where possible, the most precise text reference (i.e., phrases with qualifiers) regarding materials used in the experiment were annotated. For instance, ``carbon atoms in graphene'' was annotated as a single \textsc{Material} entity and not separately as ``carbon atoms,'' ``graphene.''

\item Any confusion in classifying scientific entities as one of four types was resolved using the following concept precedence: \textsc{Method} $>$ \textsc{Process} $>$ \textsc{Data} $>$ \textsc{Material}, where the concept appearing earlier in the list was preferred.
\end{enumerate}

After finalizing the concepts and updating the guidelines,\footnote{The annotation guidelines are released with the corpus.} the final annotation task proceeded in two phases

\begin{enumerate}
    \item In phase I, five abstracts per domain (i.e. 50 abstracts) were annotated by both annotators and the inter-annotator agreement was computed using Cohen's $\kappa$ \cite{cohen1960coefficient}. Results showed a moderate inter-annotator agreement at 0.52 $\kappa$. 
    \item Next, in phase II, one of the annotators interviewed subject specialists in each of the ten domains about the choice of concepts and her annotation decisions on their respective domain corpus.\footnote{The concepts were generally well received as generically applicable, except in Engineering where the subject specialist mentioned the need for an additional category \textsc{Tool} for phrases which our scheme captures as \textsc{Material}.} The feedback from the interviews were systematically categorized into error types and these errors were discussed by both annotators. Following these discussions, the 50 abstracts from phase I were independently reannotated. The annotators could obtain substantial overall agreement of 0.76 $\kappa$ after phase II. 
\end{enumerate}

In Table~\ref{table:2}, we report the IAA scores obtained per domain and overall. The scores show that the annotators had a substantial agreement in seven domains, while only a moderate agreement was reached in three domains, viz. \textit{Agr}, \textit{Mat}, and \textit{Ast}.

\begin{table}[!htb]
\small
\centering
\begin{tabular}{|l|r|l|l|r|r|l|r|}
\cline{1-2} \cline{4-5} \cline{7-8}
             & $\kappa$ & \multirow{5}{*}{} &              & $\kappa$ & \multirow{5}{*}{} &                                & $\kappa$   \\ \cline{1-2} \cline{4-5} \cline{7-8}
\textit{Med} & 0.94     &                   & \textit{Eng}  & 0.79     &                   & \textit{Mat}                   & 0.58       \\ \cline{1-2} \cline{4-5} \cline{7-8}
\textit{MS} & 0.90     &                   & \textit{Che}  & 0.77     &                   & \textit{Ast}                   & 0.57           \\ \cline{1-2} \cline{4-5} \cline{7-8}
\textit{CS} & 0.85     &                   & \textit{Bio} & 0.75     &                   & \multicolumn{2}{r|}{\multirow{2}{*}{Overall 0.76}} \\ \cline{1-2} \cline{4-5}
\textit{ES} & 0.81     &                   & \textit{Agr}  & 0.60     &                   & \multicolumn{2}{r|}{}           \\ \cline{1-2} \cline{4-5} \cline{7-8}
\end{tabular}
\caption{Per-domain and Overall inter-annotator agreement (Cohen's Kappa $\kappa$) for \textsc{Process}, \textsc{Method}, \textsc{Material}, and \textsc{Method} scientific concept annotation}
\label{table:2}
\end{table}

\begin{table*}[!htb]
\centering
\small
\begin{tabular}{c|rrrrrrrrrr}
                         & \textit{Ast} & \textit{Agr} & \textit{Eng} & \textit{ES} & \textit{Bio} & \textit{Med} & \textit{MS} & \textit{CS} & \textit{Che} & \textit{Mat} \\ \hline
\# Tokens overall & 3711 & 3134 & 2917 & 3065 & 2579 & 2558 & 2597 & 2383 & 1991 & 1334 \\
Avg. \# Tokens/Abstract & 382 & 333 & 303 & 321 & 273 & 274 & 282 & 253 & 217 & 140 \\
%Avg. \# Sentences/Abstract               & 14                            & 14                            & 18                            & 14                           & 13                            & 15                            & 15                           & 11                           & 10                            & 6                             \\
\# Gold scientific concept phrases                     & 791                           & 741                           & 741                           & 698                          & 649                           & 600                           & 574                          & 553                          & 483                           & 297                           \\
\# Unique gold scientific concept phrases        & 663                           & 631                           & 618                           & 633                          & 511                           & 518                           & 493                          & 482                          & 444                           & 287                           \\
\# \textsc{Process}  & 241                           & 252                           & 248                           & 243                          & 281                           & 244                           & 178                          & 220                          & 149                           & 56                            \\

\# \textsc{Method}     & 19                            & 28                            & 27                            & 9                            & 15                            & 33                            & 27                           & 66                           & 27                            & 7                             \\

\# \textsc{Material} & 296                           & 292                           & 208                           & 249                          & 291                           & 191                           & 231                          & 102                          & 188                           & 51                            \\
\# \textsc{Data}   & 235                           & 169                           & 258                           & 197                          & 62                            & 132                           & 138                          & 165                          & 119                           & 183  
\end{tabular}
\caption{The annotated corpus characteristics in terms of size and the number of scientific concept phrases}
\label{table:3}
\end{table*}

\begin{figure*}[!htb]
\includegraphics[width=\textwidth]{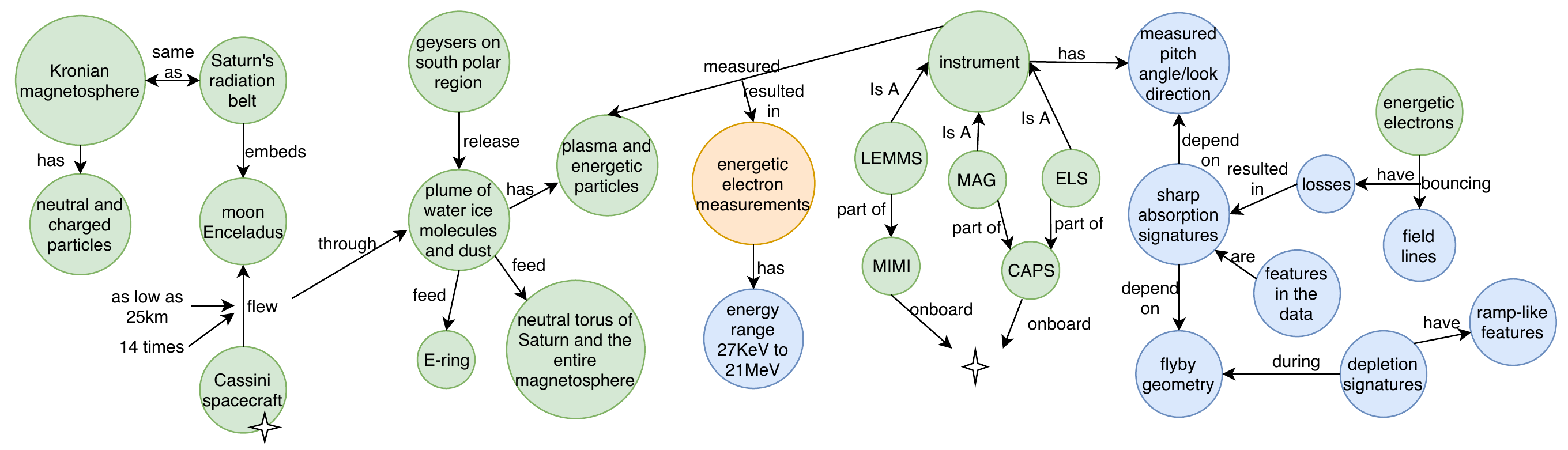}
\caption{A text graph of the abstract of the article `The Cassini Enceladus encounters 2005--2010 in the view of energetic electron measurements'~\protect\cite{krupp2012cassini}. Nodes are color-coded by concept type: orange corresponds to \textsc{Process}, green to \textsc{Material}, and blue to \textsc{Data}. No \textsc{Method} concepts were identified in this abstract.}
\label{fig:1}
\end{figure*}

\paragraph{Annotation Error Analysis}
We discuss some of the changes the interviewer annotator made in phase II after consultation with the subject experts. 

In total, 21\% of the phase I annotations were changed: \textsc{Process} accounted for a major proportion (nearly 54\%) of the changes. Considerable inconsistency was found in annotating verbs like ``increasing'', ``decreasing'', ``enhancing'', etc., as \textsc{Process} or not.  Interviews with subject experts confirmed that they were a relevant detail to the research investigation and hence should be annotated. So 61\% of the \textsc{Process} changes came from additionally annotating these verbs. \textsc{Material} was the second predominantly changed concept in phase II, accounting for 23\% of the overall changes. Nearly 32\% of the changes under \textsc{Material} came from consistently reannotating phrases about models, tools, and systems; accounting for another 22\% of its changes, where spatial locations were an essential part of the investigation such as in the \textit{Ast} and \textit{ES} domains, they were decided to be included in the phase II set as \textsc{Material}. %Additionally, nearly 5\% of the overall changes came from resolving confusion about CS entities like ``statements,'' ``these programs,'' etc., as \textsc{Data} or \textsc{Process} which were resolved to \textsc{Process} in phase II. 
Finally, there were some changes that emerged from lack of domain expertise. This was mainly in the medical domain (4.3\% of the overall changes) in resolving confusion in annotating \textsc{Process} and \textsc{Method} concept types. Most of the remaining changes were based on the treatment of conjunctive spans or lists.

Subsequently, the remaining 60 abstracts (six per domain) were annotated by one annotator. This last phase also involved reconciliation of the earlier annotated 50 abstracts to obtain a gold standard corpus.

\paragraph{Annotated Corpus Characteristics}
Table~\ref{table:3} shows our annotated corpus characteristics. 
Our corpus comprises a total of 6,127 scientific entities, including 2,112 \textsc{Process}, 258 \textsc{Method}, 2,099 \textsc{Material}, and 1,658 \textsc{Data} entities. The number of entities per abstract directly correlates with the length of the abstracts (Pearson's \textit{R} 0.97). 
Among the concepts, \textsc{Process} and \textsc{Material} directly correlate with abstract length (\textit{R} 0.8 and 0.83, respectively), while \textsc{Data} has only a slight correlation (\textit{R} 0.35) and \textsc{Method} has no correlation (\textit{R} 0.02). 
%The domains \textit{Bio}, \textit{CS}, \textit{Ast}, and \textit{Eng} have the most \textsc{Process}, \textsc{Method}, \textsc{Material}, and \textsc{Data} concepts, respectively.

In Figure~\ref{fig:1}, we show an example instance of a manually created text graph from the scientific entities in one abstract. The graph highlights that linguistic relations such as synonymy, hypernymy, meronymy, as well as OpenIE relations are poignant even between scientific entities.

\subsection{Performance Benchmark}
\input{performance_benchmarks}

In the second stage of the study, we perform word sense disambiguation and link our entities to authoritative sources. %Next, we describe these annotations on our STEM data.

\section{Scientific Entity Resolution}

Aside from the four scientific concepts facilitating a common understanding of scientific entities in a multidisciplinary setting, the fact that they are just four made the human annotation task feasible. Utilizing additional concepts would have resulted in a prohibitively expensive human annotation task. Nevertheless, there are existing datasets (particularly in the biomedical domain, e.g., GENIA~\cite{kim2003genia}) that have adopted the conceptual framework in rich domain-specific semantic ontologies. Our work, while related, is different since we target the annotation of multidisciplinary scientific entities that facilitates a low annotation entrance barrier to producing such data. This is beneficial since it enables the task to be performed in a domain-independent manner by researchers, but perhaps not crowdworkers, unless screening tests for a certain level of scientific expertise are created.

Nonetheless, we recognize that the four categories might be too limiting for real-world usage. Further, the scientific entities from stage 1 remain susceptible to subjective interpretation without additional information. Therefore, in a similar vein to adopting domain-specific ontologies, we now perform entity linking (EL) to the Wikipedia and word sense disambiguation (WSD) to Wiktionary. %In the remaining section, we first describe our annotation task, following which we offer insights into the practical performance of the task with automatic systems. %For EL, we choose Wikipedia, and for WSD, we rely on Wiktionary. We select these Wiki sources, since they are the largest and most popular collaborative resource of world and linguistic knowledge.

\begin{comment}
The other alternative would have been to select popular ontologies from each of the ten STEM domains we consider. While this alternative promises a better linking or disambiguation coverage of our entities, the search task for such ontologies is relatively harder and complicates the task where in less popular domains it might not be possible to select just one ontology, which in turn depends on how well the semantic structured space for any domain is defined. Wikipedia and Wiktionary are among the most commonly referenced primary data sources for EL and WSD tasks, respectively, and therefore we can rely on the quality of information they provide for labeling our scientific entities. In this way, we are able to ground our entities in the real-world, and as a parallel finding, discover the additional conceptual variability offered by our scientific entities over the four we consider.
\end{comment}

\subsection{Our Annotation Process}
The same pair of annotators as before were involved in this stage of the study to determine the annotation agreement. 

\subsubsection{Annotation Task Tools}
During the annotation procedure, each annotator was shown the entities, grouped by domain and file name, in Google Excel Sheet columns alongside a view of the current abstract of entities being annotated in the BRAT interface~\shortcite{stenetorp2012brat} for context information about the entities.\footnote{Excel sheets were flexible to incorporate our various ER task annotations with separate columns reserved for a unique annotation. Existing tools (e.g., INCEpTION~\cite{de2018linking}) could not be adapted to our full task.} For entity resolution, i.e. linking and disambiguation, the annotators had local installations of specific time-stamped Wikipedia\footnote{https://dumps.wikimedia.org/enwiki/20190920/} and Wiktionary\footnote{https://dumps.wikimedia.org/enwiktionary/20190920/} dumps to enable future persistent references to the links since the Wiki sources are actively revised. They queried the local dumps using the DKPro JWPL tool~\cite{zesch2008extracting} for Wikipedia and the DKPro JWKTL tool~\cite{meyer2012wiktionary} for Wiktionary, where both tools enable optimized search through the large Wiki data volume.

\subsubsection{Annotation Procedure for Entity Resolution}
\begin{figure}[!tb]
    \center{\includegraphics[width=0.6\columnwidth]
        {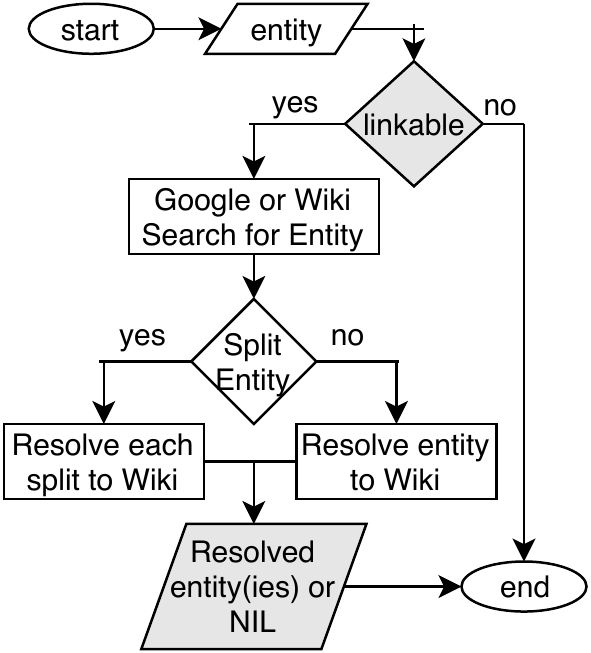}}
    \caption{Flowchart depicting our Entity Resolution annotation steps. The boxes shaded in grey represent the steps where annotator agreement scores are computed.}
    \label{fig:4}
\end{figure}

%A pilot annotation exercise was carried out to determine the procedure and guidelines on the 10 abstracts pilot dataset. This dataset, however, was now modified as it included annotations of the scientific entities. 
Through iterative pilot annotation trials on the same pilot dataset as before, the annotators delineated an ordered annotation procedure depicted in the flowchart in Figure~\ref{fig:4}. There are two main annotation phases, viz. a preprocessing phase (determining linkability, determining whether an entity is decomposable into shorter collocations), and the entity resolution phase.

The actual annotation task then proceeded, in which to compute agreement scores, the annotators worked on the same set of 50 scholarly abstracts that they had used earlier to compute the scores for the scientific entity annotations. 

\paragraph{Linkability.} In this first step, \textit{entities that conveyed a sense of scientific jargon were deemed linkable.} 

A natural question that arises, in the context of the \textit{Linkability} criteria, is: Which stage 1 annotated scientific entities were now deemed unlinkable? They were 1) \textsc{Data} entities that are numbers; 2) entities that are coreference mentions which, as isolated units, lost their precise sense (e.g., ``development''); and 3) \textsc{Process} verbs (e.g., ``decreasing'', ``reconstruct'', etc.). Still, having identified these cases, a caveat remained: except for entities of type \textsc{Data}, the remaining decisions made in this step involved a certain degree of subjectivity because, for instance, not all \textsc{Process} verbs were unlinkable (e.g., ``flooding''). Nonetheless, at the end of this step, the annotators obtained a high IAA score at 0.89 $\kappa$. From the agreement scores, we found that the \textit{Linkability} decisions could be made reliably and consistently on the data.

\paragraph{Splitting phrases into shorter collocations.} While preference was given to annotating non-compositional noun phrases as scientific entities in stage 1, consecutive occurrences of entities of the same concept type separated only by prepositions or conjunctions were merged into longer spans. As examples, consider the phrases ``geysers on south polar region,'' and ``plume of water ice molecules and dust'' in Figure~\ref{fig:1}. These phrases, respectively, can be meaningfully split as ``geysers'' and ``south polar region'' for the first example, and ``plume'', ``water ice molecules'', and ``dust'' for the second. As demonstrated in these examples, the stage 1 entities we split in this step are syntactically-flexible multi-word expressions which did not have a strict constraint on composition~\cite{sag2002multiword}. For such expressions, we query Wikipedia or Google to identify their splits judging from the number of results returned and whether, in the results, the phrases appeared in authoritative sources (e.g., as overview topics in publishing platforms such as ScienceDirect). Since search engines operate on a vast amount of data, they are a reliable source for determining phrases with a strong statistical regularity, i.e. determining collocations. 

With a focus on obtaining agreement scores for entity resolution, the annotators bypass this stage for computing independent agreement and attempted it mutually as follows. One annotator determined all splits, wherever required, first. The second annotator acted as judge by going through all the splits and proposed new splits in case of disagreement. The disagreements were discussed by both annotators and the previous steps were repeated iteratively until the dataset was uniformly split. After this stage, both annotators have the same set of entities for resolution.

\paragraph{Entity Resolution (ER) Annotation.} In this stage, the annotators resolved each entity from the previous step to encyclopedic and lexicographic knowledge bases. While, in principle, multiple knowledge sources can be leveraged, this study only examines scientific entities in the context of their Wiki-linkability. 

%\noindent{\textbf{What to resolve to?}} 
\textit{Wikipedia}, as the largest online encyclopedia (with nearly 5.9 million English articles) offers a wide coverage of real-world entities, and based on its vast community of editors with editing patterns at the rate of 1.8 edits per second, is considered a reliable source of information.\footnote{https://en.wikipedia.org/wiki/Wikipedia:Statistics} It is pervasively adopted in automatic EL tasks~\cite{bunescu2006using,mihalcea2007wikify,rao2013entity} to disambiguate the names of people, places, organizations, etc., to their real-world identities. We shift from this focus on proper names as the traditional Wikification EL purpose has been, to its, thus far, seemingly less tapped-in conceptual encyclopedic knowledge of nominal scientific entities.

\textit{Wiktionary} is the largest freely available dictionary resource. Owing to its vast community of curators, it rivals the traditional expert-curated lexicographic resource WordNet~\cite{Fellbaum1998} in terms of coverage and updates, where the latter evolves more slowly. For English, Wiktionary has nine times as many entries and at least five times as many senses compared to WordNet.\footnote{https://en.wiktionary.org/wiki/Wiktionary:Statistics versus https://wordnet.princeton.edu/documentation/wnstats7wn} As a more pertinent neologism in the context of our STEM data, consider the sense of term ``dropout'' as a method for regularizing the neural network algorithms which is already present in Wiktionary.\footnote{https://en.wiktionary.org/wiki/dropout}
While WSD has been traditionally used WordNet for its high-quality semantic network and longer prevalence in the linguistics community (c.f Navigli~\shortcite{navigli2009word} for a comprehensive survey), we adopt Wiktionary thus maintaining our focus on collaboratively curated resources. 

In WSD, entities from all parts-of-speech are enriched w.r.t. language and wordsmithing. But it excludes in-depth factual and encyclopedic information, which otherwise is contained in Wikipedia. Thus, Wikipedia and Wiktionary are viewed as largely complementary.
%has about 905,463 entries with 1,147,072 definitions, on the other hand, WordNet has 155,287 entries with 206941 word sense pairs

\paragraph{ER Annotation Task formalism.} Given a scholarly abstract $A$ comprising a set of entities $E = \{e_{1}, ... ,e_{N}\}$, the annotation goal is to produce a mapping from $E$ to a set of Wikipedia pages ($p_1,...,p_N$) and Wiktionary senses ($s_1,...,s_N$) as $R = \{(p_1,s_1), ... , (p_N,s_N)\}$. For entities without a mapping, the corresponding $p$ or $s$ refers to \textsc{Nil}.

The annotators followed comprehensive guidelines for ER including exceptions. E.g., the conjunctive phrase ``acid/alkaline phosphatase activity'' was semantically treated as the following two phrases ``acid phosphatase activity'' or ``alkaline phosphatase activity'' for EL, however, in the text it was retained as ``acid'' and ``alkaline phosphatase activity.'' Since WSD is performed over exact word-forms without assuming any semantic extension, it was not performed for ``acid.'' Annotations were also made for complex forms of reference such as meronymy (e.g., space instrument ``CAPS'' to spacecraft ``wiki:Cassini Huygens'' of which it is a part), or hypernymy (e.g., ``parents'' in ``genepool parents'' to ``wiki:Ancestor''). %For WSD, the traditional procedure of retaining the exact word-form for mapping was followed. This included symbols (e.g., units such as $^{\circ}$C, or chemical symbols as ``C'' for ``carbon'') and initialisms (e.g, ``ATP'' for ``adenosine triphosphate''). 

As a result of the annotation task, the annotators obtained 82.87\% rate of agreement in the EL task and a $\kappa$ score of 0.86 in the WSD task. Contrary to WSD expectations as a challenging linguistics task~\cite{ng1999case}, we show high agreement; this we attribute to the entities' direct scientific sense and availability in Wiktionary (e.g., ``dropout'').

Subsequently, the ER annotation for the remaining 60 abstracts (six per domain) were performed by one annotator. This last phase also involved reconciliation of the earlier annotated 50 abstracts to obtain a gold standard corpus.

\begin{table}[!tb]
\small
\begin{tabular}{l|p{2.3cm}|p{2.4cm}|p{1.3cm}}
& \textbf{EL:} total,  & \textbf{WSD:} total,    &                         \\
& \%indomain,     & \%indomain,     & \textbf{Total} (\%)     \\ 
\multicolumn{1}{l|}{}  & \multicolumn{1}{l|}{\%overall}     & \multicolumn{1}{l|}{\%overall}   & \multicolumn{1}{l}{}           \\ \hline
\multicolumn{1}{l|}{\textit{ES}}  & \multicolumn{1}{l|}{559 (80.3 / 9.9)}  & \multicolumn{1}{l|}{380 (54.6 / 6.8)} & \multicolumn{1}{l}{\textbf{696 (12.4)}} \\ 
\multicolumn{1}{l|}{\textit{Ast}} & \multicolumn{1}{l|}{\textbf{583 (87.3 / 10.4)}} & \multicolumn{1}{l|}{\textbf{477 (71.4 / 8.5)}} & \multicolumn{1}{l}{668 (11.9)} \\ 
\multicolumn{1}{l|}{\textit{Agr}}  & \multicolumn{1}{l|}{556 (83.3 / 9.9)}  & \multicolumn{1}{l|}{475 (71.2 / 8.4)} & \multicolumn{1}{l}{667 (11.9)} \\ 
\multicolumn{1}{l|}{\textit{Bio}} & \multicolumn{1}{l|}{538 (88.2 / 9.6)}  & \multicolumn{1}{l|}{319 (52.3 / 5.7)} & \multicolumn{1}{l}{610 (10.8)} \\ 
\multicolumn{1}{l|}{\textit{Eng}} & \multicolumn{1}{l|}{473 (80 / 8.4)}    & \multicolumn{1}{l|}{406 (68.7 / 7.2)} & \multicolumn{1}{l}{591 (10.5)} \\ 
\multicolumn{1}{l|}{\textit{Med}} & \multicolumn{1}{l|}{476 (82.5 / 8.5)}  & \multicolumn{1}{l|}{320 (55.5 / 5.7)} & \multicolumn{1}{l}{577 (10.3)} \\ 
\multicolumn{1}{l|}{\textit{MS}}  & \multicolumn{1}{l|}{484 (85.1 / 8.6)}  & \multicolumn{1}{l|}{377 (66.3 / 6.7)} & \multicolumn{1}{l}{569 (10.1)} \\ 
\multicolumn{1}{l|}{\textit{Che}} & \multicolumn{1}{l|}{381 (76.2 / 6.8)}  & \multicolumn{1}{l|}{304 (60.8 / 5.4)} & \multicolumn{1}{l}{500 (8.9)}  \\ 
\multicolumn{1}{l|}{\textit{CS}}  & \multicolumn{1}{l|}{391 (81 / 6.9)}    & \multicolumn{1}{l|}{285 (59 / 5.1)}   & \multicolumn{1}{l}{483 (8.6)}  \\ 
\multicolumn{1}{l|}{\textit{Mat}} & \multicolumn{1}{l|}{226 (85.3 / 4)}    & \multicolumn{1}{l|}{150 (56.6 / 2.7)} & \multicolumn{1}{l}{265 (4.7)}  \\ \hline \hline
Overall & 4667 (82.9 / - ) & 3489 (62.0 / -) & 5627 ( - ) \\
\end{tabular}
\caption{The annotated corpus characteristics in terms of its Entity Linking (EL) and Word Sense Disambiguation (WSD) annotations}
\label{table:6}
\end{table}

\begin{table}[!htb]
\centering
\small
\begin{tabular}{lrl||lr}
\multicolumn{2}{c}{\textbf{Wikipedia}}   &  & \multicolumn{2}{c}{\textbf{Wiktionary}}    \\
POS  & \%    &  & POS    & \%    \\ \hline
N    & 49.6 &  & N      & 75.3   \\
MWE  & 23.2 &  & R      & 46.2   \\
SW  & 11.8 &  & ADJ    & 16.4   \\
ADJ  & 9.7  &  & SYM    & 3.8    \\
SYM  & 2.8  &  & V      & 2.3    \\
V    & 1.3  &  & NNP    & 1.6    \\
NNP  & 1.2  &  & INIT   & 0.2    \\
R    & 0.3  &  & PREP   & 0.03   \\
INIT & 0.2  &  & PHRASE & 0.03  
\end{tabular}
\caption{Part-of-speech tag distribution in our corpus. N - Noun; MWE - multiword expression; SW - single word; ADJ - adjective; SYM - symbol; V - verb; NNP - proper noun; R - adverb; INIT - initialism; PREP - preposition}
\label{table:7}
\end{table}

\subsubsection{Annotated Corpus Characteristics}

In this stage 2 corpus, \textit{linkability} of the scientific entities was determined at 74.6\%. Of these, 61.7\% were split into shorter collocations, at 1.74 splits per split entity. Detailed statistics are presented in Table~\ref{table:6}. In the table, the domains are ranked by the total number of their linkable entities (fourth column). \textit{Ast} has the highest proportion of linked entities at 87.3\% which comprises 10.4\% of all the linked entities and disambiguated entities at 71.4\% forming 8.5\% of the overall disambiguated entities. From an EL perspective, we surmize that articles on space topics are well represented in Wikipedia. %\textit{Che} had the lowest EL proportion at 76.2\%, owing to its several chemical element formulae, however, overall \textit{Mat} has the least EL instances at 4\% which also has the least number of candidate entities. For WSD, \textit{Bio} has the least disambiguated candidates at 52.3\%, while \textit{Mat}, with the least entity candidates, has the least disambiguated instances in the entire corpus. 
For WSD, \textit{Bio}, \textit{ES}, and \textit{Med} predictably have the least proportion of disambiguated entities at 52.3\%, 54.6\%, and 55.5\%, respectively, since of all our domains these especially rely on high degree scientific jargon, while WSD generally tends to be linguistically oriented in a generic sense. As a summary, linked and disambiguated entities had a high correlation with the total linkable entities ($R$ 0.98 and 0.89, respectively).

In Table~\ref{table:7}, the ER annotation results are shown as POS tag distributions. The POS tags were obtained from Wiktionary, where entities that couldn't be disambiguated are tagged as SW (Single Word) or MWE (Multi-Word Expression). % which, respectively, comprise 11.8\% and 23.2\% of the total Wikipedia entities. 
These tags have a coarser granularity compared to the traditionally followed Penn Treebank tags with some unconventional tagging patterns (e.g., ``North Sea'' as NNP, ``in vivo'' as ADJ). From the distributions, except for nouns being the most EL and WSD instances, the rest of the table differs significantly between the two tasks in a sense reflecting the nature of the tasks. While MWE are the second highest EL instances, its corresponding PHRASE type is least represented in WSD. In contrast, while adverbs are the second highest in WSD, they are least in EL.

\subsection{Evaluation}
\input{entity-linking-eval}

We do not observe a significant impact of the long-tailed list phenomenon of unresolved entities in our data (c.f Table~\ref{table:6} only 17\% did not have EL annotations). Results on more recent publications should perhaps serve more conclusive in this respect for new concepts introduced--the abstracts in our dataset were published between 2012 and 2014.

\section{Conclusion}

The STEM-ECR v1.0 corpus of scientific abstracts offers multidisciplinary \textsc{Process}, \textsc{Method}, \textsc{Material}, and \textsc{Data} entities that are disambiguated using Wiki-based encyclopedic and lexicographic sources thus facilitating links between scientific publications and real-world knowledge (see the concepts enrichment we obtain from Wikipedia for our entities in Figure~\ref{fig:6}). We have found that these Wikipedia categories do enable a semantic enrichment of our entities over our \textit{generic} four concept formalism as \textsc{Process}, \textsc{Material}, \textsc{Method}, and \textsc{Data} (as an illustration, the top 30 Wiki categories for each of our four \textit{generic} concept types are shown in the Appendix). Further, considering the various domains in our multidisciplinary STEM corpus, notably, the inclusion of understudied domains like Mathematics, Astronomy, Earth Science, and Material Science makes our corpus particularly unique w.r.t. the investigation of their scientific entities. This is a step toward exploring domain independence in scientific IE. Our corpus can be leveraged for machine learning experiments in several settings: as a vital active-learning test-bed for curating more varied entity representations~\cite{brack2020domain}; to explore domain-independence versus domain-dependence aspects in scientific IE; for EL and WSD extensions to other ontologies or lexicographic sources; and as a knowledge resource to train a reading machine (such as PIKES~\cite{corcoglioniti2016frame} or FRED~\cite{gangemi2017semantic}) that generate more knowledge from massive streams of interdisciplinary scientific articles. We plan to extend this corpus with relations to enable building knowledge representation models such as knowledge graphs in a domain-independent manner. 

%\citelanguageresource{speecon}

\section{Acknowledgements}
We thank the anonymous reviewers for their comments and suggestions. We also thank the subject specialists at TIB for their helpful feedback in the first part of this study. This work was co-funded by the European Research Council for the project ScienceGRAPH (Grant agreement ID: 819536) and by the TIB Leibniz Information Centre for Science and Technology.

\section*{Appendix: Supplemental Material}

%\textbf{A.1 Scientific Entity Extraction}
\textbf{A.1. Proportion of the \textit{Generic} Scientific Entities}

To offer better insights to our STEM corpus for its scientific entity annotations made in part 1, in Figure~\ref{fig:3} below, we visually depict the proportion of \textsc{Process}, \textsc{Method}, \textsc{Material}, and \textsc{Data} entities per domain.

\begin{figure}[!h]
    \center{\includegraphics[width=\linewidth]
        {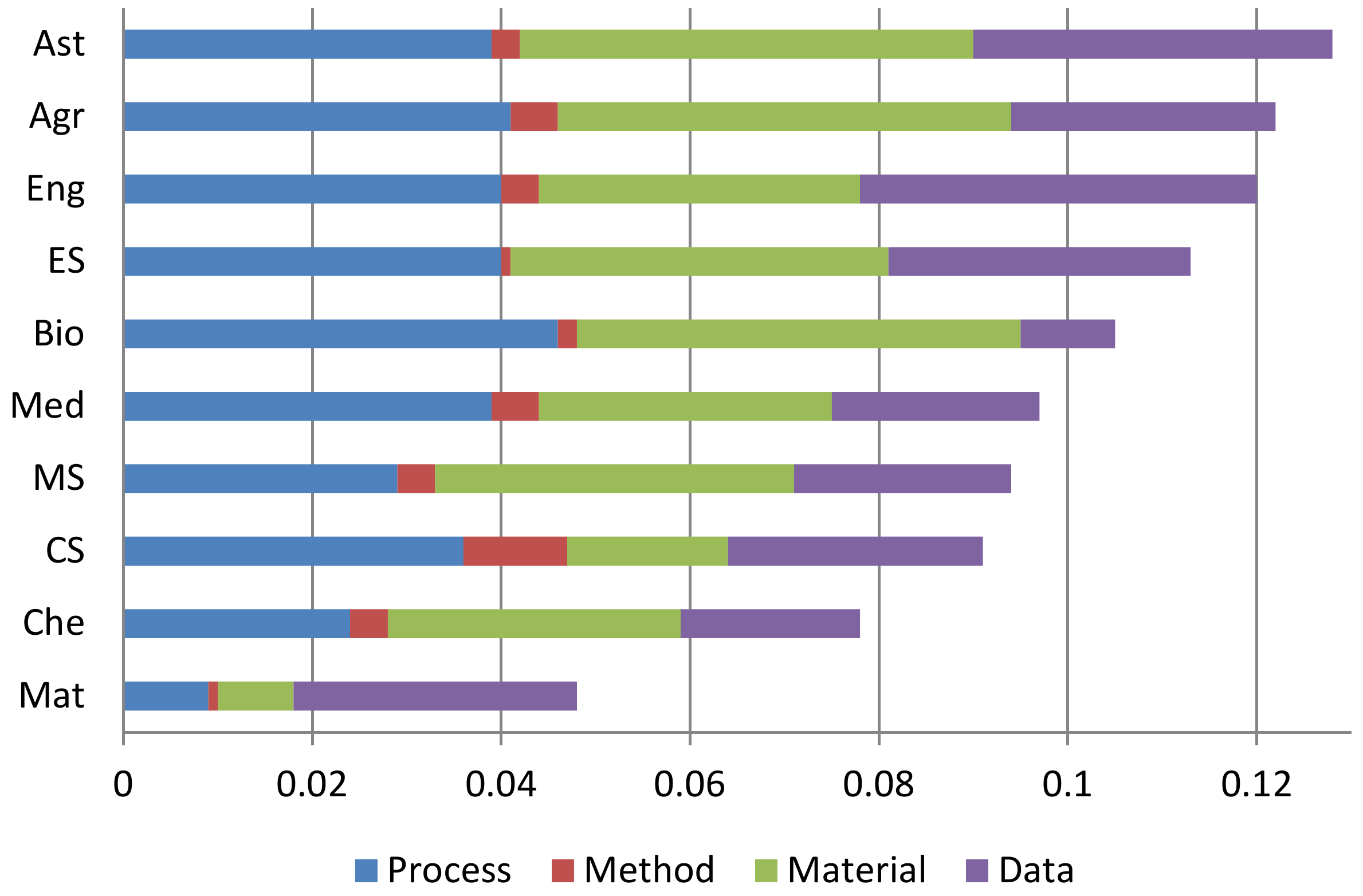}}
    \caption{Percentage proportion of scientific entities in our STEM corpus per domain and per \textit{generic} concept type}
    \label{fig:3}
\end{figure}

The Figure serves a complementary view to our corpus compared with the dataset statistics shown in Table~\ref{table:3}. It shows that the \textit{Ast} domain has the highest proportion of scientific entities overall. On the other hand, per \textit{generic} type, \textit{Bio} has the most \textsc{Process} entities, \textit{CS} has the most \textsc{Method} entities, \textit{Ast} has the most \textsc{Material} closely followed by \textit{Agr}, and \textit{Eng} has the most \textsc{Data}.

\textbf{A.2. Cohen's $\kappa$ Computation\footnote{Cohen's $\kappa$ scores are computed using the following script: https://github.com/jorgearanda/kappa-stats} Setup in Section 4.1.2}

\textbf{Linkability.} Given the stage 1 scientific entities, the annotators could make one of two decisions: a) an entity is linkable; or b) an entity is unlinkable. These decisions were assigned numeric indexes, i.e. 1 for decision (a) and -1 for decision (b) and can take on one of four possible combinations based on the two annotators decisions: (1,1), (1,-1), (-1,1), and (-1,-1). The $\kappa$ scores were then computed on this data representation.

\textbf{WSD Agreement.} 
In order to compute the WSD agreement, the Wiktionary structure for organizing the words needed to be taken into account. It's structure is as follows. Each word in the Wiktionary lexicographic resource is categorized based on etymology, and within each etymological category, by the various part-of-speech tags the word can take. Finally, within each POS type, is a gloss list where each gloss corresponds to a unique word sense. 

Given the above-mentioned Wiktionary structure, the initial setup for the blind WSD annotation task entailed that the annotators were given the same reference POS tags within an etymology for the split single-word entities in the corpus.\footnote{In separate experiments for determining etymological and POS agreement between the annotators, we found they had an insignificant disagreement (i.e. in less than 15 instances in approximately over the 2000 they annotated.)} 
Next, as data to compute $\kappa$ scores, each annotator-assigned gloss sense was given a numeric index and agreement was computed based on matches or non-matches between indexes.

\textbf{A.3. Per-domain Inter-annotator Agreement for Entity Resolution}

To supplement the overall Inter-Annotator Agreement (IAA) scores reported in Section 4.1.2 `Entity Resolution (ER) Annotation' for the EL and WSD tasks, in Table~\ref{table:10} below, we additionally report the IAA scores for our ER tasks (i.e., EL and WSD) per domain in the STEM-ECR corpus. First, considering the domains where the highest ER agreement scores were obtained. For EL, the IAA score was highest in the \textit{MS} domain. While for WSD, the IAA score was highest in the \textit{Bio} domain. Next, considering the domains where the agreement was least for the two tasks. We found the the EL agreement was least for \textit{CS} and the WSD agreement was least for \textit{Mat}. In the case of low EL agreement, it can be attributed to two main cases: only one of the annotators found a link; or the annotators linked to related pages on the same theme as the entity (e.g., wiki:Rule-based\_modeling versus wiki:Rule-based\_machine\_learning for ``rule-based system''). And in the case of the low WSD agreement obtained on \textit{Mat}, we see that owing to broad terms like ``set,'' ``matrix,'' ``groups,'' etc., in the domain which could be disambiguated to more than one Wiktionary sense correctly, the IAA agreement was low.

\begin{table}[!htb]
\centering
\begin{tabular}{l|r|r|}
    & Wikipedia $roa$ & Wiktionary $\kappa$ \\ \hline
\textit{Agr} & 83.88           & 0.88                \\
\textit{Ast} & 79.8            & 0.85                \\
\textit{Bio} & 84.73           & \textbf{0.93}       \\
\textit{Che} & 86.83           & 0.85                \\
\textit{CS}  & 72.58           & 0.86                \\
\textit{ES}  & 82.17           & 0.83                \\
\textit{Eng} & 78.83           & 0.84                \\
\textit{MS}  & \textbf{88.24}  & 0.83                \\
\textit{Mat} & 87.37           & 0.81                \\
\textit{Med} & 87.89           & 0.87               
\end{tabular}
\caption{Per-domain inter-annotator agreement for Entity Linking to Wikipedia in terms of the percentage rate of agreement score ($roa$) and of Word Sense Disambiguation to Wiktionary in terms of the Cohen's kappa score ($\kappa$)}
\label{table:10}
\end{table}

\textbf{A.4. Babelfy's Precision ($P$) and Recall ($R$) Computation for Entity Resolution in Figure~\ref{fig:5}}

For the $P$ and $R$ scores reported in Figure~\ref{fig:5}, the true positives (\textit{TP}), false negatives (\textit{FN}), true negatives (\textit{TN}), and false positives (\textit{FP}) were computed as follows:

\textbf{\textit{TP}} = human-annotated entities that have a EL/WSD match with Babelfy results (for \textsc{Nil}, a match is considered as no result from the automatic system); 

\textbf{\textit{FN}} = human-annotated entities that have no EL/WSD match with Babelfy results; 

\textbf{\textit{TN}} = spurious Babelfy-created strings as entities that do not have a EL/WSD result; and 

\textbf{\textit{FP}} = spurious Babelfy-created entities that have a EL/WSD result.

\begin{table*}[!tb]
\centering
\begin{tabular}{|p{3cm}p{4.2cm}|p{4.2cm}p{3cm}|}
\multicolumn{2}{c}{\textbf{\textsc{Process}}} & \multicolumn{2}{c}{\textbf{\textsc{Method}}} \\
PhysicalQuantities                            & ElectricAndMagneticFields-InMatter & NumericalDifferentialEquations                & SystemsBiology       \\
Mutation                                      & Mechanics                         & ComputationalFluidDynamics                    & Petrology            \\
ConceptsInPhysics                             & GenesOnHumanChromosome            & ScientificTechniques                          & FiniteDifferences    \\
MathematicalAnd-QuantitativeMethods           & Software                          & Measurement                                   & Spectroscopy         \\
Electron                                      & ChemicalElements                  & LandManagement                                & Teletraffic          \\
MolecularEvolution                            & QuantumElectrodynamics            & TransportLayerProtocols                       & DataTransmission     \\
DevelopmentalBiology                          & Aerodynamics                      & Metabolism                                    & ProteinStructure     \\
VectorCalculus                                & Spintronics                       & EnzymeInhibitors                              & Optics               \\
RadiationHealthEffects                        & AllAccuracyDisputes               & NetworkPerformance                            & MolecularBiology     \\
EvolutionaryBiology                           & Oceanography                      & Research                                      & OperationsResearch   \\
Rates                                         & Dementia                          & MathematicalAnd-QuantitativeMethods           & AllAccuracyDisputes  \\
ChargeCarriers                                & Audiology                         & TechnologicalFailures                         & Stratigraphy         \\
Leptons                                       & CognitiveDisorders                & ModelingAndSimulation                         & SoilImprovers        \\
PsychiatricDiagnosis                          & WaterPollution                    & MathematicalOptimization                      & Solid-stateChemistry \\
FluidDynamics                                 & StructuralProteins                & MedicinalChemistry                            & Polarization(Waves) 
\end{tabular}
\caption{Top 30 Wikipedia categories pertaining to \textsc{Process} and \textsc{Method} scientific entities in the STEM-ECR corpus}
\label{table:11}
\end{table*}

\begin{table*}[!tb]
\centering
\begin{tabular}{|p{4.2cm}p{4cm}|p{3.0cm}p{3.2cm}|}
\multicolumn{2}{c}{\textbf{\textsc{Material}}} & \multicolumn{2}{c}{\textbf{\textsc{Data}}}  \\
StemCells   & NaturalResources   & PhysicalQuantities  & OrdersOfMagnitude \\
CellBiology & FluidDynamics & ConceptsInPhysics   & UnitsOfTime  \\
InducedStemCells & Matter & SIBaseQuantities & FluidDynamics \\
Biotechnology  & GranularityOfMaterials  & ContinuumMechanics  & Geochronology  \\
DevelopmentalBiology & ChemboxHavingGHSData & SIDerivedUnits & Density \\
Cloning & Polymers & PyrotechnicFuels & PartsOfADay \\
HorticultureAndGardening                 & CeramicMaterials        & ChemicalElements    & UnitsOfLength           \\
BiologyAndPharmacologyOf-ChemicalElements & Adhesives               & StateFunctions      & MathematicalConstants   \\
LandManagement                           & SpacePlasmas            & ChemicalProperties  & Length                  \\
HumanEyeAnatomy                          & TaxaNamedByCarlLinnaeus & Research            & ImperialUnits           \\
ChemicalElements                         & ChelatingAgents         & UnitsOfTemperature  & Symptoms                \\
Soil                                     & ChemboxHavingDSDData    & Metrology           & Calendars               \\
MedicalTerminology                       & Humans                  & Size                & UnitsOfLength           \\
ChemboxImageSizeSet                      & Neurons                 & Measurement         & UnitsOfPlaneAngle       \\
NaturalMaterials                         & E-numberAdditives       & ChemboxImageSizeSet & SolarCalendars         
\end{tabular}
\caption{Top 30 Wikipedia categories pertaining to \textsc{Material} and \textsc{Data} scientific entities in the STEM-ECR corpus}
\label{table:12}
\end{table*}

\FloatBarrier

\textbf{A.5. Top 30 Wikipedia Categories for \textsc{Process}, \textsc{Method,} \textsc{Material}, and \textsc{Data}}

In part 1 of the study, we categorized the scientific entities by our four \textit{generic} concept formalism, comprising \textsc{Process}, \textsc{Method}, \textsc{Material}, and \textsc{Data}. Linking the entities to Wikipedia further enables their broadened categorization. While in Figure~\ref{fig:6} is depicted the rich set of Wikipedia categories obtained overall, here, in Tables~\ref{table:11} and \ref{table:12}, we show the top 30 Wikipedia categories for the scientific entities by their four concept types. we observe the most of the Wikipedia categories pertinently broaden the semantic expressivity of each of our four concepts. Further that in each type, they are diverse reflecting the underlying data domains in our corpus. As examples, consider the Wikipedia categories for the \textsc{Data} scientific entities: ``SIBaseQuantities'' category over the entity ``Kelvin'' in \textit{Che}; ``FluidDynamics'' in \textit{Eng} and \textit{MS} domains; and ``SolarCalendars'' in the \textit{Ast} domain.

%\cleardoublepage

% \nocite{*}
\section{Bibliographical References}\label{reference}
%\label{main:ref}

\bibliographystyle{lrec}
\bibliography{lrec}

%\section{Language Resource References}
%\label{lr:ref}
%\bibliographystylelanguageresource{lrec}
%\bibliographylanguageresource{languageresource}

\end{document}

%% file: introduction.tex
Recently, an increasing number of research efforts are geared toward adopting Knowledge Graphs (KG) for modeling scholarly publications~\cite{ammar2018construction,Jaradeh2019ORKG}. They advocate for such advanced semantic machine-interpretability, via KGs, of the publications to enable their more intelligent automated processing (e.g., in search applications). To this end, KGs leveraged in academic and industrial settings fostering better search already serve as case-in-point, albeit so far only over commonsense world knowledge (consider the DBpedia~\cite{auer2007dbpedia} and Google~\cite{kg} KGs as two prominent examples among others).
%For instance, to gain a survey of existing innovations on a problem, the researcher still needs to scour through dozens of papers online over having to later assimilate the knowledge contained, a situation which is only more complex in a multidisciplinary setting. 

To represent scholarly publications as KGs, from an Information Extraction (IE) perspective, and scientific IE, in particular, extracting scientific entities %\footnote{Essentially, they are scientific terms, however, we refer to them as entities to posit them as valid knowledge graph node candidates.} 
from scholarly publications becomes a vital task to address since entities lie at the core of KGs. While these scientific entities are essentially scientific terms, they are referred to with the broader notion of \textit{entities} in the rest of the paper to posit them as valid knowledge graph node candidates.
As an IE task, the extraction of scientific entities is being increasingly studied in the Natural Language Processing (NLP) community--the SemEval series itself has so far seen four tasks organized~\cite{kim2010semeval,moro2015semeval,augenstein2017semeval,gabor2018semeval}. However, very little of this work~\cite{handschuh2014acl} has been done in the tradition of corpus linguistics and none, so far, in the broad multidisciplinary setting of Science.

In this vein, with the aim of providing a platform for benchmarking methods of scholarly article processing, in 2017 Elsevier Labs released an open access corpus of publications across Science, Technology, Engineering, and Medicine (STEM),\footnote{https://github.com/elsevierlabs/OA-STM-Corpus} thus providing a shared test bed for multidisciplinary scientific IE research. The corpus is based on the 10 most prolific STEM disciplines, viz. Agriculture (\textit{Agr}), Astronomy (\textit{Ast}), Biology (\textit{Bio}), Chemistry (\textit{Che}), Computer Science (\textit{CS}), Earth Science (\textit{ES}), Engineering (\textit{Eng}), Materials Science (\textit{MS}), and Mathematics (\textit{Mat}). The study presented subsequently in this paper was performed on scientific abstracts in this STEM corpus.

In general, the challenges associated with scientific IE are seen to be greater than for, say newswire articles, in part because of the expected domain expertise required to annotate resources for machine learning making the annotation task costly, in turn limiting resources. Since, at present, a study of human agreement of annotating scientific entities across STEM is lacking, \textit{consequently, we have little understanding about at least a certain set of scientific concepts that are generically applicable and the extent to which they can be decided without domain knowledge.} Quantitatively evaluating this based on a set of \textit{generic} scientific concepts is the first goal of this paper. We refer to this problem as the \textit{generic scientific concept extraction} task, for which our research question is: Can a generic formalism of scientific concepts offer human annotators who possess sufficient scientific competence, the ability to reliably annotate a multidisciplinary scholarly corpus with scientific entities? Examining this question, leads us in turn to make the observation that the problem of such otherwise costly data creation can be made accessible to a wider group of people with the minimal involvement of domain specialists. 

In the second part of this study, with the goal of retaining accurate entities, we disambiguate our annotated entities via encyclopedic links and lexicographic senses. Specifically, we carry out Entity Resolution (ER)~\cite{getoor2012entity} via Entity Linking (EL) and Word Sense Disambiguation (WSD) annotations. We adopt their respective standard task definitions for our scientific entities, where EL~\cite{rao2013entity,usbeck2015gerbil} entails linking entities to the most suitable entry in a knowledge base, and WSD~\cite{navigli2009word} involves explicitly assigning meanings to single-word and multi-word occurrences within text. Performing joint EL and WSD keeps us on course with the current ``best of two worlds'' trend in ER research considering their seamless integration as complementary semantic information units for disambiguating the entities~\cite{hovy2013collaboratively} to obtain boosted performance in their automatic disambiguation~\cite{Moroetal:14tacl}. In addition, we note that without ER, our entity annotations can seem somewhat subjective on the one hand. While on the other hand, they might be limiting for real-world usage. With ER, we make the scientific entities locatable in the real world thus removing to a certain degree their subjective impression, while facilitating their categorization (conceptualization) according to the metadata in the knowledge base, without restricting users to our four concepts.

Further, the introduction of this STEM corpus facilitates the testing of the ER approaches in a unique setting: \textit{to that of being restricted within the single broad domain of Science, while still being multidisciplinary}, thus, in a sense, validating the automatic systems for their semantic adaptability. Training algorithms on such a corpus entails switching senses of the the same word. E.g., ``the Cloud'' in \textit{CS} should be resolved to a technological sense, versus in \textit{Ast} where it takes the common interpretation of the mass of water vapor we see in the sky; or ``neural networks'' which in \textit{CS} are an algorithm versus in \textit{Bio} or \textit{Med} which refer to the brain. %It would require removing any NER bias of the automatic systems. E.g., the \textit{ES} entity ``Drake Passage'' gets linked by open-domain systems as one of two NER senses, viz. as the English explorer ``Francis Drake'' or the hip-hop musician `Drake,'' as opposed to the geographical sense as the body of water to which it refers.
It would require the seemingly evident shift from common interpretations of terms, though not always. E.g., ``power'' in \textit{Mat} refers to exponentiation, which, otherwise in a common sense, takes on a human social interpretation; or even a common word as ``ring'' in \textit{Mat} is an algebraic structure versus the jewelry common interpretation; or ``subject'' in \textit{Med} as research participant rather than an academic discipline. However, we note that our corpus not only enables evaluating systems for their semantic adaptability in a joint fashion for both EL and WSD tasks, it can also be leveraged for designing approaches attempting either one of the tasks, on the premise of their dichotomy, offering still novel insights in this regard owing to its multidisciplinary nature.

Finally, in this study, for both parts of the STEM data, we present benchmark results with insights on respective model performances: 1) for entity recognition, we train a BERT-based neural model~\cite{DevlinCLT19} on our data of scientific entities; and 2) for entity resolution, we manually evaluate Babelfy~\cite{Moroetal:14tacl} for EL and WSD of the scientific entities in our corpus. 

In summary of our contributions, we: 1) release a novel multidisciplinary STEM corpus of scientific entities under a \textit{generic} conceptual formalism bridging STEM scientific domains; where the entities are further enriched with EL and WSD annotations, thus, both disambiguating their scientific sense and grounding them in the real world, thereby enabling additional semantic extensions of the entities; and 2) provide benchmark performances from state-of-the-art systems on our STEM data. For further research, the STEM-ECR v1.0 corpus can be downloaded at the following link: https://doi.org/10.25835/0017546 (ISLRN 749-555-840-571-2). %http://www.islrn.org/resources/749-555-840-571-2/. %https://gitlab.com/TIBHannover/orkg/orkg-nlp/tree/master/STEM-ECR-v1.0.

The remaining paper is organized in two main parts: 1) annotating scientific entities in a STEM setting; and 2) linking and sense disambiguation of the selected entities. We begin with a discussion on related work to ours.

%% file: related_work.tex
Early initiatives in semantically structuring scholarly publications focused on sentences as the basic unit of analysis using data from one or two scientific domains.
To this end, ontologies and vocabularies were created~\cite{teufel1999annotation,Soldatova2006AnOO,constantin2016document,pertsas2017scholarly}, 
corpora were annotated~\cite{Liakata2010CorporaFT,Fisas2016AMA}, 
and machine learning methods were applied~\cite{liakata2012automatic}.
Recently, scientific IE has targeted search technology, thus newer corpora have been annotated at the phrasal unit of information with three or six types of scientific concepts~\cite{handschuh2014acl,augenstein2017semeval,Luan2018MultiTaskIO} facilitating machine learning system development~\cite{Ammar2017TheAS,luan2017scientific,beltagy2019scibert}.
But, again these corpora are from a single or at most three domains.

In this work, while we continue to address scientific IE at the phrasal unit, unlike existing work, by explicitly situating our task in the wide-ranging STEM scholarly communication setting, we do not impose a domain restriction on the task. Additionally, we enrich our entities with semantic EL and WSD annotations.

On this related note, the SemEval 2015 Task 13~\cite{moro2015semeval} corpus exists with joint EL and WSD annotations for entities. However, theirs isn't a scientific IE task since the genre of text they annotate does not comprise scholarly publications. Further, their annotations differs from ours based on the knowledge source they use, where ours is identical to theirs on the encyclopedic front but a subset of theirs on the lexical knowledge since their lexical knowledge is derived from at least three different sources. Nevertheless, restricting ourselves to just one lexicographic source, lets us test its coverage uniformly.

It was recently noted that the guidelines clarifying the inclusion or exclusion of entities take on an implicit inclusion criteria which only becomes apparent when noting the differences between the entities in datasets that share a common creation objective~\cite{rosales-mendez-etal-2019-fine}. In this work, we adopt a fine-grained annotation strategy following a safer maximal inclusion route.

\begin{comment}
``species rich,bare,soil'' ``bare'' refers to ``unplanted'' as a scientific entity and ``unconvered'' in a lexical sense.
``selection'' refers to ``Natural Selection'' as an entity and ``The process or act of selecting.'' as a word sense.
\end{comment}

%% file: performance_benchmarks.tex
In this section, we report the performance of a state-of-the-art neural system for extracting scientific entities from our STEM corpus. Specifically, we leverage SciBERT~\cite{beltagy2019scibert}, a pretrained language model based on BERT~\cite{DevlinCLT19} but trained on a large corpus of scientific text, as frozen embedding features in a NER task-specific neural model comprising BiLSTM layers and a CRF-based sequence tag decoder~\cite{Ma2016EndtoendSL}.%\footnote{Our training of this system on ScienceIE~\shortcite{augenstein2017semeval} and SciERC~\shortcite{Luan2018MultiTaskIO}, respectively, after hyperparameter tuning on their dev sets achieves 47.5 $F1$ and 66.0 $F1$ test set scores demonstrating state-of-the-art performance in contrast to the latest reported 46.6 $F1$~\shortcite{luan2017scientific} and 64.2 $F1$~\shortcite{Luan2018MultiTaskIO}.} 

\paragraph{Models.}

Using the above architecture implemented in the AllenNLP framework~\cite{gardner2018allennlp}, we train one model with data from all domains combined. We call this the \textit{domain-independent} scientific entity extraction system. To test robust models, we performed five-fold cross validation experiments. In each fold experiment, we trained a model on 8 abstracts per domain (i.e. 80 abstracts; $>$4000 entities), tuned hyperparameters on 1 abstract per domain (i.e. 10 abstracts; $>$500 entities), and tested on the remaining 2 abstracts (i.e. 20 abstracts; $>$1000 entities) ensuring that data splits were not identical between folds.

\paragraph{Results and Discussion.}

%Table~\ref{table:4} shows an overview of the \textit{domain-independent} scientific entity extraction system results. 
This system achieves 64.3\% precision, 66.7\% recall, and 65.5\% overall task $F1$ at a stable 0.0140 standard deviation per fold. For it, \textsc{Material} was the easiest concept to extract at 71\% $F1$, followed by \textsc{Process} at 66.8\%, followed by \textsc{Data} at 59.8\%, and \textsc{Method} the hardest at 43\% $F1$. Of note, \textsc{Method} is also the most underrepresented in our corpus which could in part account for its poor extraction performance.

%In general, in both settings, we have obtained stable per-fold classifiers since our systems show low $F1$ standard deviation ($\sigma$) across folds. The \textit{domain-independent} classifier has $\sigma$ of 0.0140; and $\sigma$ of the respective \textit{domain-specific} classifiers when tested on their respective domains in increasing order are: \textit{Eng} (0.035); \textit{CS} (0.046); \textit{ES} (0.056); \textit{Ast} (0.057); \textit{Agr} (0.059); \textit{Mat} (0.061); \textit{Med} (0.071); \textit{Che} (0.081); \textit{Bio} (0.09); \textit{MS} (0.113). This could in part be attributed to our use of features in the form of SciBERT embeddings for machine learning that were pretrained on scientific articles.

Further, in Table~\ref{table:20}, we report the performance of this system per domain. It demonstrates highest performance on the \textit{Eng} and \textit{Bio} domains at 0.71 $F1$ and shows significantly lower performance on \textit{Mat} at 0.48 $F1$. Again, compared to the remaining 9 domains in our corpus, \textit{Mat} has the least annotated entities, which we attribute as the cause for the low extraction performance. We hypothesize that more training instances could improve performance. For more detailed machine learning results, we refer the reader to our related work~\cite{brack2020domain}. %where we have reported performances also from \textit{domain-specific} trained classifiers among other analysis.

\begin{table}[!htb]
\small
\centering
\begin{tabular}{|l|r|l|l|r|r|l|r|}
\cline{1-2} \cline{4-5} \cline{7-8}
 & $F1$ & \multirow{5}{*}{} &   & $F1$ & \multirow{5}{*}{} & & $F1$   \\ \cline{1-2} \cline{4-5} \cline{7-8}
\textit{Eng} & 0.71 & & \textit{Ast}  & 0.66  & & \textit{Med} & 0.61 \\ \cline{1-2} \cline{4-5} \cline{7-8}
\textit{Bio} & 0.71 & & \textit{CS} & 0.65 & & \textit{Mat} & 0.48 \\ \cline{1-2} \cline{4-5} \cline{7-8}
\textit{MS} & 0.69 & & \textit{Che} & 0.64 & &  \multicolumn{2}{r|}{\multirow{2}{*}{Overall 0.65}} \\ \cline{1-2} \cline{4-5}
\textit{Agr} & 0.68 & & \textit{ES} & 0.63 & & \multicolumn{2}{r|}{} \\ \cline{1-2} \cline{4-5} \cline{7-8}
\end{tabular}
\caption{The \textit{domain-independent} scientific entity extraction system results per-domain}
\label{table:20}
\end{table}

%% file: entity-linking-eval.tex
We evaluate Babelfy~\cite{Moroetal:2014iswc}, a domain-agnostic, graph-based unified approach to EL and WSD. Since it is an unsupervised system, we do not retrain it on our data. We evaluate it for %a novel setting, i.e.
joint EL and WSD performance on \textit{multidisciplinary} scientific entities. 

\begin{figure}[!tb]
    \center{\includegraphics[width=\linewidth]
        {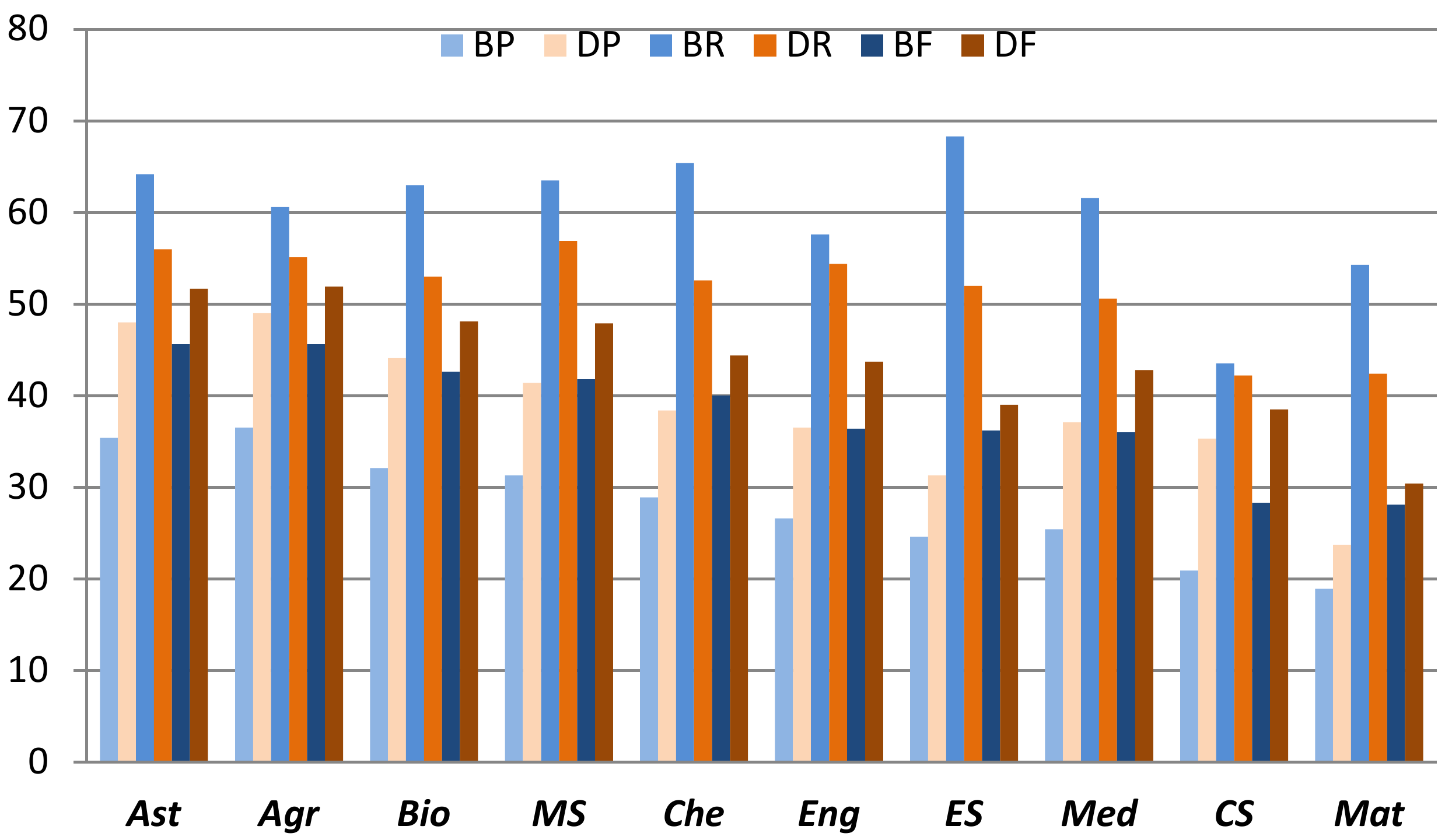}}
    \caption{Percentage $P$, $R$, and $F1$ per domain of the Babelfy system for disambiguating to BabelNet (as shaded blue bars BP, BR, and BF) and for linking to DBpedia (as shaded orange bars DP, DR, DF)}
    \label{fig:5}
\end{figure}

\textbf{Evaluation 1.1: Splitting phrases into substrings} The first step of an end-to-end ER system like Babelfy is spotting the candidate text fragments to resolve. Babelfy domain-agnostically follows a substring matching heuristic which either at sentence-unit or phrase-unit has the same operating principle, i.e. fragmenting the input into non-overlapping longest matching linkable spans and recursively splitting each span to smaller content-word parts. %Thus, we can directly use Babelfy to emulate the human annotators at splitting stage 1 entities to shorter parts. 

Babelfy splits 96.2\% of entities from stage 1 as opposed to 61.7\% of them split by humans. Further, the annotators split entities at 1.74 parts per entity. On the other hand, Babelfy split them at 2.46 parts per entity (2.16 for non-overlapping fragments). Thus, evidently our human annotators have taken a conservative stance since they focused only on scientific entities, differing from the Babelfy approximately-all linkable domain-agnostic fragmentations (e.g., given ``pearl millet'', the annotators link it as it is, whereas Babelfy links it for ``pearl'' and ``pearl millet'').

In addition, Babelfy has a recall of 63.2\% at precision 37.8\% of the human-split entities. Thus, despite Babelfy's approximately-all linkable fragmentation, it shows a relatively low recall for our entities, since we note that Babelfy in most cases operates on at least partial lexical matches whereas several of our entities required inference decisions.

\textbf{Evaluation 1.2: Entity Resolution} In Figure~\ref{fig:5}, we depict the Babelfy ER performance in terms of the classical precision ($P$), recall ($R$) and $F1$ scores.\footnote{We provide details about our calculations in the Appendix.} 

\textbf{EL versus WSD}
For all domains, EL is better than WSD in line with the claim that WSD is generally deemed a harder linguistics task than EL.\footnote{For EL, DBpedia and Wikipedia are equivalent resources; for WSD, BabelNet draws from more than one source including Wiktionary which we consider.} The highest EL score is obtained on \textit{Agr} (at 51.9\% $F1$), and the highest WSD score is on \textit{Ast} and \textit{Agr} (at 45.6\% $F1$). 

\textbf{Most ambiguous resolution domain(s)}
\textit{CS} and \textit{Mat} were the most ambiguous resolution domains for Babelfy. This finding is consistent with that reported in the Semeval 2013 shared task~\cite{moro2015semeval} for these two domains which are common in our datasets, albeit on different genre of text.

Finally, we did not find any consistent trend in the automatic resolution results to infer the influence of domain-specific terms on the performance.

\textbf{Evaluation 2: EL and WSD baseline system accuracies for only the linked split entities}

\begin{figure}[!htb]
    \center{\includegraphics[width=\linewidth]
        {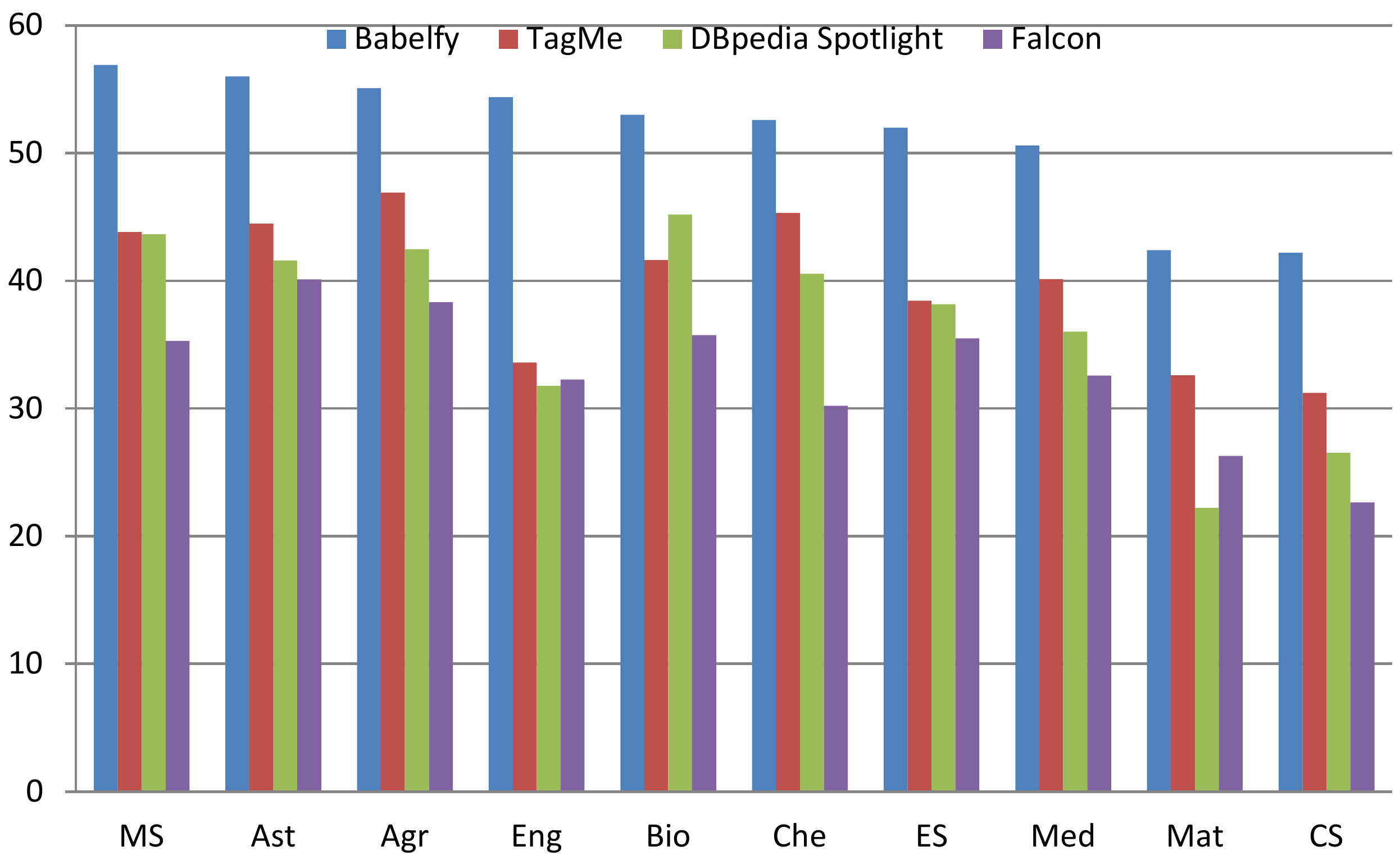}}
    \caption{Percentage recall per domain of four popular online Entity Linking systems, viz. Babelfy~\protect\shortcite{Moroetal:2014iswc}, TagMe~\protect\cite{ferragina2010tagme}, DBpedia Spotlight~\protect\cite{mendes2011dbpedia}, and Falcon~\protect\cite{sakor2019old}}
    \label{fig:7}
\end{figure}

For EL, we select multiple baselines including the exact title match heuristic and four popular online systems that demonstrate state-of-the-art performance on non-scientific, open domain data. The baseline scores are: exact title match heuristic at 37.8\% accuracy, and from the four automated systems, viz. Babelfy~\shortcite{Moroetal:2014iswc} to DBpedia at 52.6\%, TagMe~\cite{ferragina2010tagme} at 40.5\%, DBpedia Spotlight~\cite{mendes2011dbpedia} at 38.1\%, and Falcon~\cite{sakor2019old} at 33.8\%. Further in Figure~\ref{fig:7}, we show the performances of these systems obtained per domain. Of all four systems, Babelfy consistently has highest recall of the linked entities owing to its maximal linking objective, i.e. linking longest phrase matches and their sub-parts. Per-domain, Babelfy has highest recall on \textit{MS}, TagME on \textit{Agr}, DBpedia Spotlight on \textit{Bio}, and Falcon on \textit{Ast}, respectively.

For WSD, our baseline is just the first sense heuristic, since it is seen hard-to-beat for this task. It demonstrates 68.8\% disambiguation accuracy.

\begin{figure}[!tb]
\small
    \center{\includegraphics[width=0.8\linewidth]
        {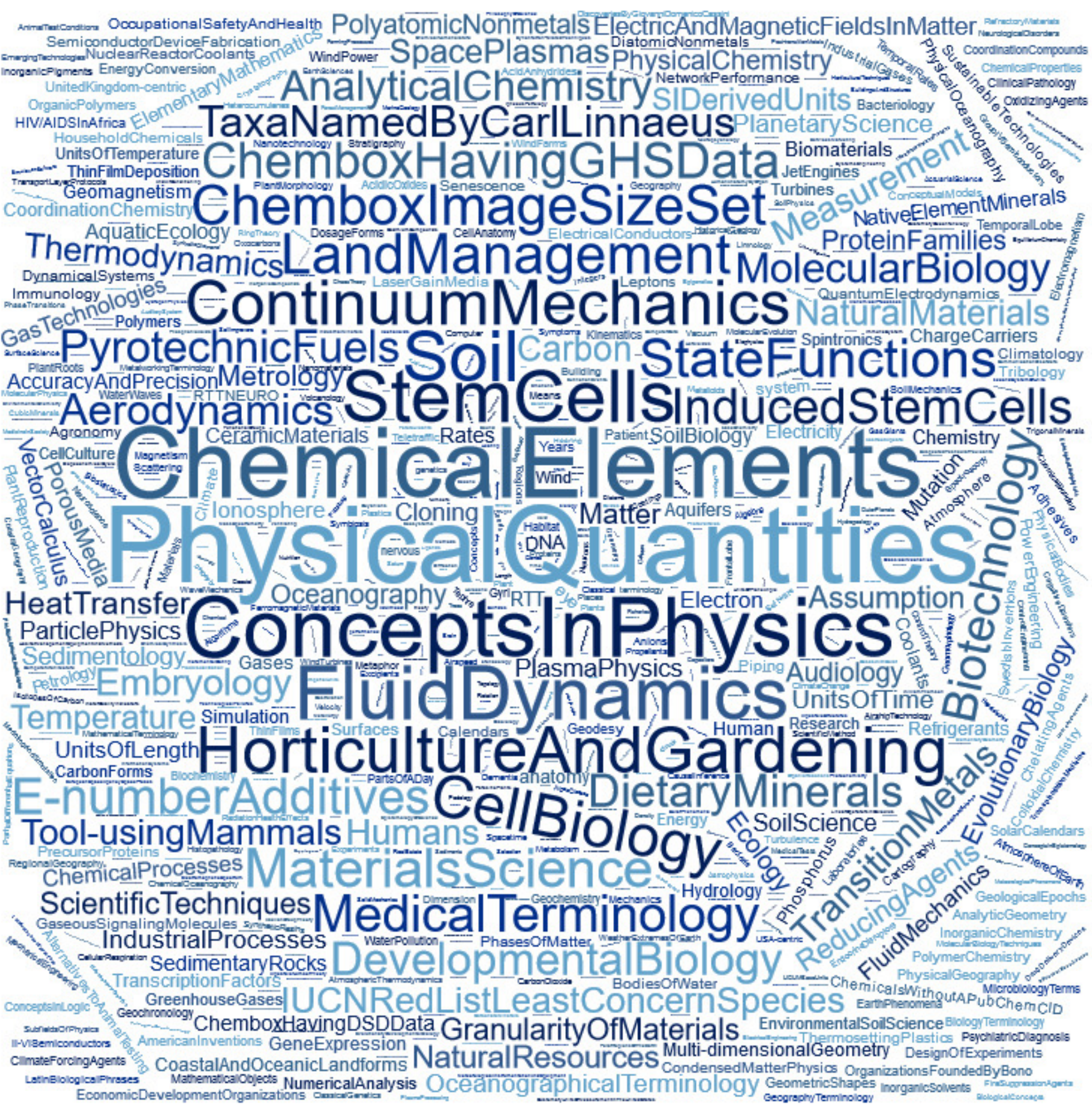}}
    \vspace{-1em}
    \caption{Wikipedia categories cloud on scientific entities}
    \label{fig:6}
\end{figure}